# Large domain wall resistance in self-organised manganite film

L. Granja[a)]

*Departamento de Física, Comisión Nacional de Energía Atómica, Gral. Paz 1499 (1650) San Martín, Buenos Aires, Argentina.*

L. E. Hueso[b)], D. Sánchez, J. L. Prieto[c)]

*Department of Materials Science, University of Cambridge, Pembroke Street, Cambridge CB2 3QZ, UK.*

P. Levy

*Departamento de Física, Comisión Nacional de Energía Atómica, Gral. Paz 1499 (1650) San Martín, Buenos Aires, Argentina.*

N. D. Mathur[d)]

*Department of Materials Science, University of Cambridge, Pembroke Street, Cambridge CB2 3QZ, UK.*

The electrical resistance of magnetic domain walls in ferromagnetic metallic manganites can be enhanced to ~$10^{-12}$ $\Omega m^2$ by patterning nanoconstrictions [J. Appl. Phys. **89**, 6955 (2001)]. We show equally large enhancements in a phase-separated $La_{0.60}Ca_{0.40}MnO_3$ manganite film without recourse to nanopatterning. The domain walls were measured in the current-perpendicular-to-the-plane (CPP) geometry between ferromagnetic metallic $La_{0.70}Ca_{0.30}MnO_3$ electrodes patterned like magnetic tunnel junctions.

---

[a)] Also at Department of Materials Science, University of Cambridge, Pembroke Street, Cambridge CB2 3QZ, UK.
[b)] Also at Spintronic Devices Division, ISMN-CNR, Via P. Gobetti 101, 40129 Bologna, Italy.
[c)] Also at Intituto de Sistemas Optoelectrónicos y Microtecnología (ISOM) Universidad Politécnica de Madrid (UPM) Avda. Complutense s/n - Madrid 28040 – Spain.
[d)] Corresponding author (N.D.M.): ndm12@cam.ac.uk

Magnetic tunnel junctions[1] traditionally comprise electrodes of a common magnetic metal such as cobalt, separated by an ultra thin barrier such as $Al_2O_3$. The use[2-4] of ferromagnetic metallic (FMM) manganite electrodes of $La_{0.70}Sr_{0.30}MnO_3$ or $La_{0.70}Ca_{0.30}MnO_3$ with an epitaxial insulating barrier of $SrTiO_3$ or $NdGaO_3$ (NGO) has proved particularly successful because of the high electrode spin polarisation[5] and the clean crystalline interfaces.[6]

Here we investigate manganite devices that are similar to the standard manganite tunnel junctions[2,3] in that we use FMM $La_{0.70}Ca_{0.30}MnO_3$ (LCMO30) electrodes, but different because the ultra-thin barrier is replaced with 20 nm of $La_{0.60}Ca_{0.40}MnO_3$ (LCMO40). Even though bulk[7] LCMO40 is regarded to be a FMM, LCMO40 films appear to be phase separated[8] (FMM+insulator) with a FMM volume fraction of <50%.[9] Indeed, phase separation is a hot topic in the manganites[10,11], and the FMM phase is often seen to coexist with insulating phases[12-14] which leads, for example, to non-volatile memory effects.[15] Here, our study of phase separated LCMO40 is motivated by the fact that it should form a weak magnetic link between LCMO30 electrodes and thus permit the formation of domain walls for electrical transport studies.

A manganite trilayer of Au/LCMO30/LCMO40/LCMO30//NGO was grown with nominal layer thicknesses 50 nm (bottom) and 20 nm (middle and top), by pulsed laser deposition using conditions for LCMO described elsewhere.[9] X-ray diffraction was used to confirm the epitaxy of the heterostructure (Figure 1, inset). The 50 nm gold capping layer was grown in-situ (room temperature, $~10^{-6}$ Torr, 2.5 J.cm$^{-2}$, 1 Hz) in order to improve contact resistance.

Mesa devices with areas in the range A=5×5 μm$^2$ to 19×19 μm$^2$ were fabricated from the trilayer film using optical lithography, Ar ion milling, and sputtering to deposit silica insulation and gold contacts. The devices were fabricated in a manner equivalent to the standard tunnel junctions reported in [3], apart from the above-mentioned replacement of the tunnel barrier with 20 nm of LCMO40 and the use of in-situ Au. In this Letter, we show electrical transport data for the 5×5 μm² mesa only. Four-terminal CPP resistance measurements were performed in a He closed cycle cryostat, as a function of both temperature, and magnetic field applied parallel to the orthorhombic [100] film easy[16] axis. As with standard tunnel junctions, these are not true four-terminal measurements since the top contact is shared between current and voltage leads in view of the fact that it is physically small. For all devices, current-voltage characteristics were found to be linear at all measurement temperatures (20-300 K) and fields (±0.5 T), but only out to 150 μA, corresponding to the small voltage of ~3 mV, in order to avoid undue heating in the low resistance CPP mesa geometry. Magnetic characterization of the unpatterned trilayer film was made using a vibrating sample magnetometer (VSM).

In Fig. 1 we present the zero-field device resistance as a function of temperature, which displays a distinct metal-insulator transition peaking just below 200 K. This corresponds well to the metal insulator transition for LCMO40 films on NGO[9]. The LCMO30 transition (at 265 K in films on NGO[3]) is not seen, which is reasonable given that the resistivity and therefore the resistance of the LCMO40 at 265 K is relatively high[9]. The arrowed feature at ~80 K is attributed to the contact resistance associated with the top electrode because the feature height varies according to the metallic element employed and the processing details.[17,18] For instance, we have observed that Ar ion milling prior to Au deposition significantly increases the feature height.

Fig. 2 shows device resistance as a function of applied magnetic field at 25 K after zero-field cooling. The observed two-state low-field switching is highly reproducible between field sweeps and cooling runs, with switching fields $\left|B_{c1}^{R}\right|$ = (7±2) mT and $\left|B_{c2}^{R}\right|$ = (75±2) mT. In addition, a high-field magnetoresistance MR ~20%/T with limited hysteresis is

apparent. In Fig. 3 we present a minor hysteresis loop which permits a zero-field determination of the high and low resistance states of the device, which differ by (0.7±0.2) Ω. Both configurations were found to be stable during measurements performed over 90 minutes. The two-state switching (Figs. 2&3) is only present below ~50 K, but any variation in its magnitude with temperature is beyond the resolution of our experiments.

Results were qualitatively similar for all the mesa sizes that we studied. We identify $\left|B_{c1}^{R}\right|$ with bottom electrode switching because it was found to be independent of mesa size. By contrast, $\left|B_{c2}^{R}\right|$ falls with increasing mesa size with $\left|B_{c2}^{R}\right|$ = (55±2) mT for our largest 19×19 μm² mesa. Resistance–area (RA) product values were independent of area for $A \leq$ 10×10 μm², indicating that the low-temperature performance of our smallest 5×5 μm² device (Figs. 2&3) is not influenced by MR artefacts[19] associated with inhomogeneous current paths[20]. This is reasonable given that at low temperatures the resistivity of LCMO30 films (~100 μΩ.cm)[16] is 20 times smaller than the resistivity of LCMO40 films (~2000 μΩ.cm)[9], because this ratio translates directly into the ratio of the resistances of the top LCMO30 electrode and the underlying LCMO40 layer, each of which occupies the same volume in the mesas.

The high-field MR of ~20%/T at 25 K (Fig. 2) is too large to be associated with the intrinsic behaviour of the LCMO30 and LCMO40 layers. This is because epitaxial LCMO30 only shows[21] a few %/T at low temperatures, and LCMO40 alone would show[22] 20%/T at 50 K, but here it is series with the rest of the device including the small top contact whose resistance is high. Indeed, this top contact dominates device resistance at 25 K, which limits our low-field MR to 3.5%. Based on preliminary studies that we have performed[23] we suggest that the observed high-field MR is primarily due to the Ar ion milling, which affects every LCMO layer in the device, but not the in-situ top contact.

Fig. 4 displays 10 K easy-axis magnetic hysteresis measurements of an unpatterned trilayer that was codeposited with the film used for our devices. The hysteresis loop shows two abrupt jumps at $|B_{c1}^M|$ = (4.0±0.5) mT and $|B_{c2}^M|$ = (8.9±0.5) mT, which we suggest correspond to the switching of the two LCMO30 layers. We assign $|B_{c1}^M|$ and $|B_{c2}^M|$ to the switching fields of the bottom and top layers respectively, by comparison with our previous work on similar trilayers where magnetic switching corresponded clearly to layer thicknesses[24]. However, in this case the jumps associated with the electrodes switching are smaller than expected (1.0 μ$_B$/Mn bottom layer, 2.4 μ$_B$/Mn top layer). This is reasonable given that single-domain switching is not guaranteed in films of millimeter side-length.

Combining the magnetic and electrical data, we find that the bottom LCMO30 layer is characterized by $|B_{c1}^R| > |B_{c1}^M|$, and the top layer is characterized by $|B_{c2}^R| >> |B_{c2}^M|$. The differences in these electrically and magnetically measured switching fields arise as a consequence of the patterning that was required for the electrical measurements. The small increase for the bottom layer is attributed to the top of it being damaged from overmilling during mesa definition. The larger increase for the top layer is attributed to a combination of the demagnetizing field in the mesa, and milling-induced edge roughness anisotropy as seen in manganite tunnel junctions.[25]

The electrical switching seen in Figs. 2&3 is attributed to the creation and destruction of magnetic domain walls, presumably in the LCMO40 interlayer which acts as a weak magnetic link between the LCMO30 electrodes. (Note that epitaxial films of LCMO30 alone do not show low-field MR switching.[26]) We suggest that the ~50% FMM volume fraction[9] of LCMO40 halves the effective conducting area of the 5×5 μm$^2$ mesa, such that the data in Fig. 3 represents a domain wall RA~10$^{-11}$ Ωm$^2$. Note that we are assuming that the length scale of the phase separation is smaller than the side of the mesa. This is reasonable given sub-micron phase separation in the bulk[12], and indeed it is very plausible given the observation of equivalent RA values in several devices (up to and including 10×10 μm² mesas).

The domain wall resistance we measure here (RA~$10^{-11}$ Ωm²) is eight orders of magnitude larger than the values reported for standard magnetic metals such as cobalt (RA ~ $4.1\times10^{-19} - 8.7\times10^{-19}$ Ωm²).[27] As previously suggested,[11,28] this could be due to the presence of additional structure, such as the formation of a mesoscopic insulating phase at the wall centre. Indeed, manganites are known to display mesoscopic[10] and unexpected phases.[29] However, our current-voltage characteristics do not reveal the nature of the insulating wall, given that we could only apply ~3 mV, as explained earlier.

Domain wall studies in LCMO30 thin film lateral devices, which were motivated by the suggestion of mesoscopic wall centre phases, recorded a domain wall resistance of RA~$10^{-13}$ Ωm².[28] We suggest that the two order of magnitude improvement presented here could arise for several reasons. Firstly, the electronic doping in our LCMO40 interlayer is more extreme, such that this composition shows phase separation effects[8,9] that could produce a mesoscopic structure in domain wall centres. Secondly, the strain fields in our vertical structures are likely to be different from those in lateral devices. Thirdly, the CPP measurement geometry here differs from the current-in-the-plane (CIP) geometry used for the lateral structures.[28] Fourthly, the domain walls here are likely to be constrained in at least one dimension because (a) the 20 nm LCMO40 layer is thinner than the natural domain wall width of 38 nm for LCMO30,[30] and (b) conduction through the LCMO40 layer may take place via filamentary pathways.[8,24] Note that geometrical constraints alone can be responsible for large domain wall resistances,[31] and could explain recently observed values[32] of RA~$2.5\times10^{-13}$ Ωm², in a device of the large bandwidth manganite $La_{2/3}Sr_{1/3}MnO_3$ where phase separation is less likely, especially if the active region of each constriction, designed to trap domain walls, has a reduced radius of curvature[31] due to degradation from milling.

The experiments presented here exploit both half-metallicity and phase separation in the manganites. The half-metallic character of the LCMO30 electrodes permits efficient spin injection and analysis. The phase-separated character[8] of the LCMO40 interlayer evidently weakens the magnetic coupling between the LCMO30 electrodes, and permits the possibility of mesoscopic self-organised domain wall structures. Note that an equally large but irreproducible $RA \sim 10^{-11}$ $\Omega m^2$ was previously obtained[33] by nanopatterning a FMM manganite to create domain walls at nanoconstrictions. By contrast, we reproducibly achieve this value within continuous crystal lattices. It is hard to understand the size of the effects that we observe if nanostructures do not form. One may regard the formation of nanostructures within continuous crystal lattices to represent a Third Way[34] of approaching nanotechnology, beyond the standard top-down and bottom-up approaches.

We thank M.G. Blamire for a critical reading of an earlier version of this manuscript, amd M. Bibes for helpful comments. This work was supported by the EU Alban Program (L.G.), the UK EPRSC, The Royal Society (N.D.M.), an EU Marie Curie Fellowship (L.E.H.) and ANPCYT PICT No. 03-13517 (L.G. and P.L.).

[1] M. Jullière, *Phys. Lett. A* **54**, 225 (1975).

[2] J. Z. Sun, L. Krusin-Elbaum, P. R. Duncombe, A. Gupta and R. B. Laibowitz, Appl. Phys. Lett. **70**, 1769 (1997).

[3] M. -H Jo, N. D. Mathur, N. K. Todd and M. G. Blamire, Phys. Rev. B **6**1, R14905 (2000).

[4] M. Bowen, M. Bibes, A. Barthélémy, J.-P. Contour, A. Anane, Y. Lemaître and A. Fert Appl. Phys. Lett. **82**, 233 (2003).

[5] J. -H. Park, E. Vescovo, H. -J. Kim, C. Kwon, R. Ramesh and T. Venkatesan, Nature **392**, 794 (1998).

[6] W. H. Butler & A. Gupta, Nature Mater. **3,** 845–847 (2004).

[7] S.-W. Cheong and H. Y. Hwang, in *Colossal Magnetoresistive Oxides*, edited by Y. Tokura (Gordon and Breach, Amsterdam, 2000).

[8] E.C. Israel et al., manuscript in preparation.

[9] D. Sánchez, L.E. Hueso, L. Granja, P. Levy and N.D. Mathur, Appl. Phys. Lett. **89**, 142509 (2006).

[10] N. Mathur and P. Littlewood, Physics Today **56**, 25 (2003).

[11] N. D. Mathur and P. B. Littlewood, Solid State Comm. **119**, 271 (2001).

[12] M. Uehara, S. Mori, C. H. Chen, and S. -W. Cheong, Nature (London) **399**, 560 (1999)

[13] J. C. Loudon, N. D. Mathur and P. A. Midgley, Nature (London) **420**, 797 (2002).

[14] P. Schiffer, A. P. Ramirez, W. Bao and S. -W. Cheong, Phys. Rev. Lett **75**, 3336 (1995).

[15] P. Levy, F. Parisi, L. Granja, E. Indelicato and G. Polla, Phys. Rev. Lett. **89**, 137001 (2002).

[16] N. D. Mathur, M. -H. Jo, J. E. Evetts and M. G. Blamire, J. Appl. Phys. **89**, 3388 (2001).

[17] A. Plecenik, K. Fröhlich, J. P. Espinós, J. P. Holgado, A. Halabica, M. Pripko and A. Gilabert, Appl. Phys. Lett. **81**, 859 (2002).

[18] Granja et al., manuscript in preparation.

[19]R. J. M. van de Veerdonk, J. Nowak, R. Meservey, J. S. Moodera and W. J. M. de Jorge Appl. Phys. Lett. **71**, 2839 (1997).

[20]J. S. Moodera, L. R. Kinder, J. Nowak, P. LeClair, and R. Meservey, Appl. Phys. Lett. **69**, 708 (1996).

[21]A. Gupta, G. Q. Gong, G. Xiao, P. R. Dumcombe, P. Lecoeur, P. Trouilloud, Y. Y. Wang, V. P. Dravid and J. Z. Sun. . Phys. Rev. B **54**, R15629 (1996).

[22]J. C. Chapman, PhD thesis, University of Cambridge (2005).

[23]L. Granja, PhD thesis, General San Martín National University (2007).

[24]L. E. Hueso, L. Granja, P. Levy,and N. D. Mathur, J. Appl. Phys. **100**, 23903 (2006).

25. M. -H. Jo, N. D. Mathur and M. G. Blamire, Appl. Phys. Lett. **80**, 2722 (2002).

[26]M.G. Blamire, B.-S. Teo, J.H. Durrell, N.D. Mathur, Z.H. Barber, J.L. MacManus-Driscoll, L.F. Cohen and J.E. Evetts. J Magn Magn Mater **191,** 359 (1999).

[27]U. Rüdiger, J. Yu, L. Thomas, S. S. P. Parkin and A. D. Kent, Phys. Rev. B **59**, 11914 (1999).

[28]N. D. Mathur, P. B. Littlewood, N. K. Todd, S. P. Isaac, B. -S. Teo, D. -J. Kang, E. J. Tarte, Z. H. Barber, J. E. Evetts and M. G. Blamire, J. Appl. Phys. **86**, 6287 (1999).

[29]J.C. Loudon, N.D. Mathur and P.A. Midgley, Nature **420**, 797 (2002).

[30]S. J. Lloyd, N. D. Mathur, J. C. Loudon and P. A. Midgley, Phys. Rev. B **64**, 172407 (2001).

[31]P. Bruno, Phys. Rev. Lett. **83**, 2425 (1999).

[32]T. Arnal, A.V. Khvalokvskii, M. Bibes, Ph. Lecoeur, A.-M. Haghiri-Gosnet, B. Mercey, cond-mat/0610338.

[33]J. Wolfman, A. M. Haghiri-Gosnet, B. Raveau, C. Vieu, E. Cambril, A. Cornette and H. Launois, J. Appl. Phys. **89**, 6955 (2001).

[34]N. Mathur and P. Littlewood, Nature Materials **3,** 207 (2004).

FIGURE CAPTIONS

**Figure 1.** Zero-field CPP device resistance versus temperature, at a measurement current of 10 µA. The large metal-insulator transition is attributed to the LCMO40 interlayer. The arrowed feature is attributed to degraded LCMO30 beneath the top electrode.[17,19] Inset: High-resolution x-ray scan of the trilayer film used to make the device. Peaks indexed in the pseudo-cubic notation.

**Figure 2.** Major loops showing CPP device resistance versus applied magnetic field, at 25 K after zero-field cooling. The field was oriented parallel to the orthorhombic [100] easy axis[15] of the LCMO30 layers, and swept in ±550 mT. The two-state switching is attributed to the LCMO30 electrode magnetization configurations indicated.

**Figure 3.** Minor loop corresponding to the major loop from Figure 2. The low and high resistance states in zero field are attributed to the presence of domain walls in percolating pathways in LCMO40. These domain walls represent self-organised nanostructures. (Absolute resistance values subject to run-to-run thermal drift. Initial field =-550 mT.)

**Figure 4.** Magnetization versus applied magnetic field for an unpatterned trilayer of LCMO30(20 nm)/LCMO40(20 nm)/LCMO30(50 nm)/NGO, measured at 10 K along the in-plane orthorhombic [100] direction, and corrected for the paramagnetic NGO substrate. The phase separated LCMO40 middle layer weakens the ferromagnetic coupling between the LCMO30 top and bottom layers, permitting parallel and antiparallel magnetic states.

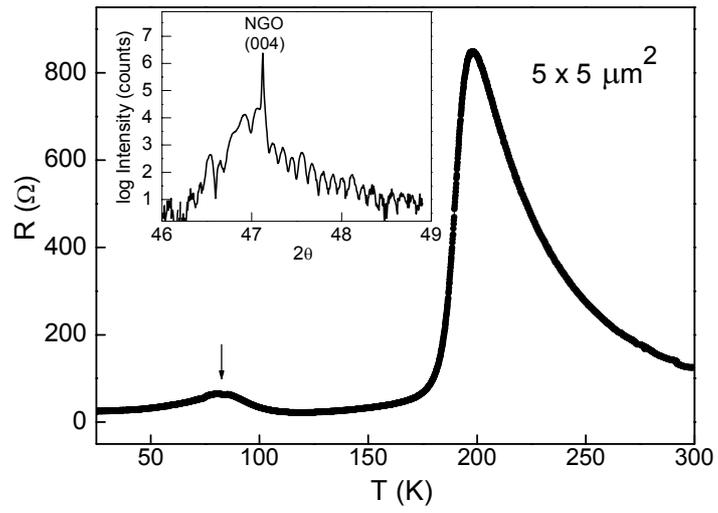

**Figure 1**

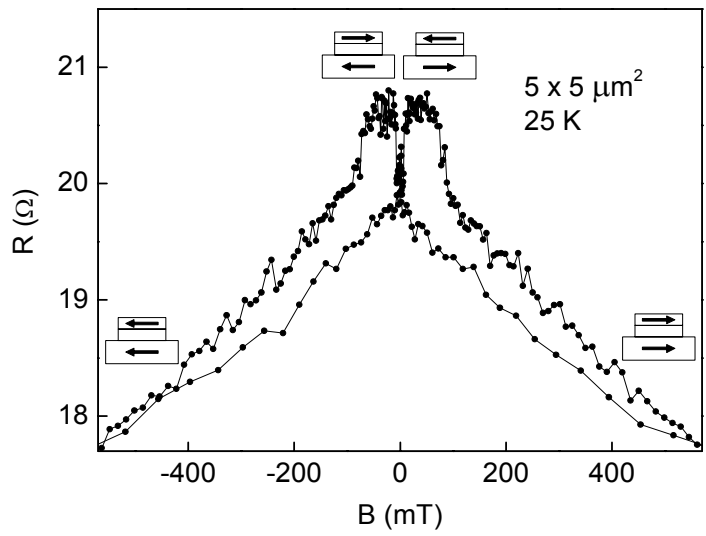

**Figure 2**

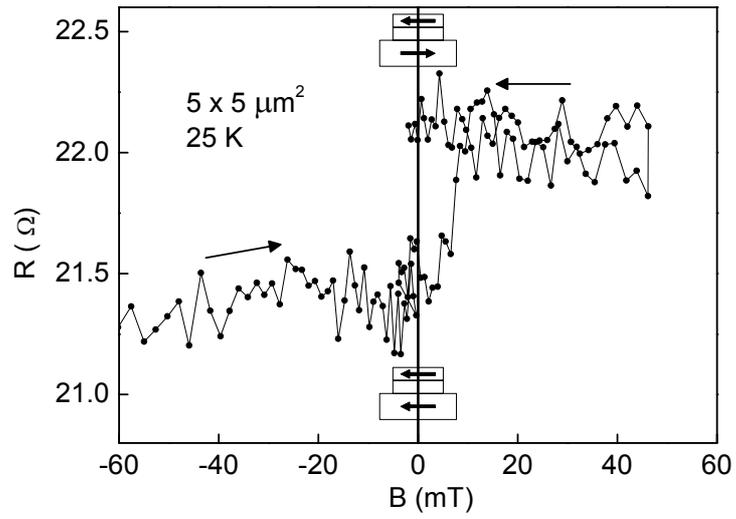

**Figure 3**

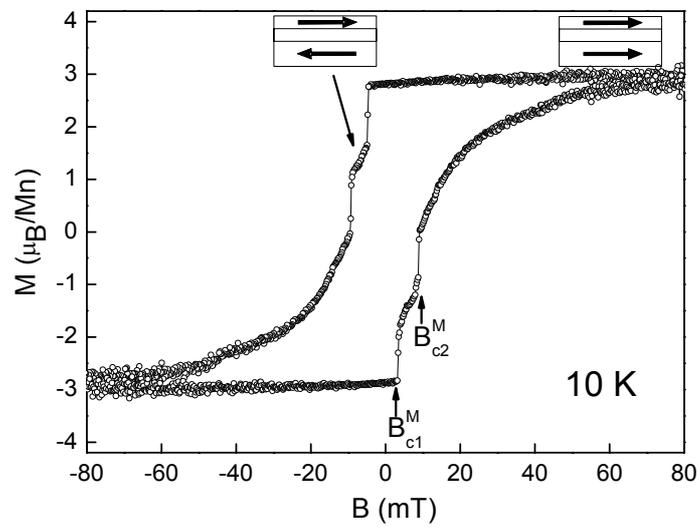

**Figure 4**